\documentclass[letter]{aastex631}
\usepackage[utf8]{inputenc}

\usepackage{multirow}
\usepackage{hhline}

\usepackage{enumerate}
\usepackage{textcomp}
\usepackage{amsmath,amssymb}
\usepackage{graphicx}
\usepackage{color}
\definecolor{orange}{rgb}{1,0.5,0}
\definecolor{purple}{rgb}{0.5,0,1}
\definecolor{darkgreen}{rgb}{0,0.5,0}
\definecolor{brown}{rgb}{0.5,0,0}
\usepackage{amsmath}

\newcommand{\beq}{\begin{equation}}
\newcommand{\eeq}{\end{equation}}
\newcommand{\ba}{\begin{array}}
\newcommand{\ea}{\end{array}}

\begin{document}

\title{Gamma-ray burst interaction with the circumburst medium: The CBM phase of GRBs}

\author[0000-0001-8667-0889]{Asaf Pe'er}
\affiliation{Bar-Ilan University, Ramat-Gan 5290002, Israel}

\author[0000-0001-8667-0889]{Felix Ryde}
\affiliation{Department of Physics, KTH Royal Institute of Technology and The Oskar Klein Centre \\ 
SE-10691 Stockholm, Sweden}

\begin{abstract}
Progenitor stars of long gamma-ray bursts (GRBs) could be surrounded by a significant and complex nebula structure lying at a parsec scale distance.  After the initial release of energy from the GRB jet, the jet will interact with this nebula environment. We show here that for a large, plausible parameter space region, the interaction between the jet blastwave and the wind termination (reverse) shock is expected to be weak, and may be associated with a precursor emission. As the jet blast wave encounters the contact discontinuity separating the shocked wind and the shocked interstellar medium, we find that a bright flash of synchrotron emission from the newly-formed reverse shock is produced. This flash is expected  to be observed at around $\sim 100$~s after the initial explosion and precursor. 
Such a delayed emission thus constitutes a circumburst medium  (CBM) phase in a GRB, having a physically distinct origin from the preceding prompt phase and the succeeding afterglow phase.  The CBM phase emission may thus provide a natural explanation to bursts observed to have a precursor followed by an intense, synchrotron-dominated main episode that is found in a substantial minority, $\sim 10$~\% of GRBs. A correct identification of the emission phase is thus required to infer the properties of the flow and of the immediate environment around GRB progenitors.  
\end{abstract}


\section{Introduction}

In recent years, it has become evident that long duration ($\geq 2$~s) gamma-ray bursts (GRBs) are associated with the deaths of massive stars, presumably originating from their core collapse \citep{Woosley93, LE93, MW99, MacFadyen+01, ZWM03, YL05}.
In this collapsar scenario, the progenitor of long GRBs is a rapidly rotating, low-metalicity, massive star. 
Evidence for this scenario include (i) an observed association between GRB and energetic core-collapse SN \citep[][for a very partial list]{Galama+98, Bloom+99, Reeves+02, Hjorth+03, WB06, Chornock+10, Starling+11, Cano+17, Wang+17, Izzo+19, Kann+19, Melandri+22, Fulton+23}, as well as (ii) the association of GRBs with star-forming galaxies and star-forming regions within galaxies \citep{Pac98, Wijers+98, Fruchter+99, TRB02, SGL09, Castro+10}; for recent reviews, see, e.g., \citet{Peer15, KZ15, Levan+16}.
Possible candidates are Wolf-Rayet stars \citep{WH06,Chevalier2000,  LMY10}, which are stars that have lost their hydrogen envelopes, leaving a small, compact core that allows the GRB jet to break out from the star. Wolf-Rayet stars have a rapid mass-loss and, in many cases, a complex circumstellar environment \citep{ Eldridge+06, CT2010}. For instance, in the local universe, around a third of Wolf-Rayet (WR) stars are observed to have a distinct ring nebulae \citep{JH65, marston97}. In some cases, even multiple rings are observed \citep{Marston95, DAB13, Lau+22}.  The ring nebulae are thought to have been caused either by the massive winds sweeping up the circumstellar medium, or by instabilities in the outer envelope, giving rise to giant eruption events \citep{Chu81, Crowther07}. They lie at a distances of a few parsec from the central star \citep{SB10}, with some lying as close as 0.7 pc \citep{Cohen+05, Sirianni+98}. In addition, the wind-blown bubbles form a low-density cavity with densities as low as $10^{-3}$ -- $10^{-2}$ cm$^{-3}$ \citep{TG13} and, in some cases, the nebula itself is located inside a cavity in the interstellar medium \citep{Vam+16}.

The GRB phenomenon is typically divided into two phases, the prompt phase and the afterglow phase. The prompt phase is the initial, triggered, emission which is predominantly observed in the gamma-rays, and consists of variable light curves, often with distinct pulses. Its origin is assumed to be from within the GRB jet, either from the photosphere or as a result of internal shocks \citep[e.g., ][]{Meszaros2006} or, alternatively, magnetic energy dissipation \citep{McKinney11}. The light-curve of the afterglow, on the other hand,  is much smoother, dominating in the X-rays and at lower photon energies. It is typically attributed to the self-similar evolution of the blastwave  interaction with the circumburst medium. 
This dichotomy with two GRB phases, however,  neglects the possible contribution of the interaction of the jet with the immediate surroundings of the progenitor \citep[e.g., ][]{Lazzati02, Mirabal2003, Chevalier2004, Eldridge+06, CT2010}.  In particular, since massive stars are expected to form complex and significant circumstellar environments, or nebulae, during their final stages in their evolution, the interaction between the jet and this circumburst medium (CBM) 
might form a distinct emission phase in GRBs,  which therefore could be denoted the "CBM phase". Such a phase would be distinct in origin to both the typical prompt and the afterglow phases. The initial prompt phase emission might therefore be followed by a CBM phase emission resulting from the interaction between the jet and the ring nebula, with a delay that depends on the size of the nebuale and the jet Lorentz factor. Since the rings are at most mildly relativistic and the jet is highly relativistic, the contrast in Lorentz factors will be large, leading to efficient kinetic energy dissipation. This, in turn, will results in significant synchrotron emission as the jet crosses the nebulae, and thus intense gamma-ray emission.

Indeed, many GRB light-curves have an initial precursor which is  separated from the main emission \citep[e.g., ][]{Murakami91, Koshut95, Vanderspek04, Piro05}, with prominent and bright examples given by GRB160625B \citep{Zhang+18, Ravasio+18} and GRB160821A \citep{Sharma+19}. It was shown by \citet{Lazzati05} and \citet{Zhu15}, that a small, but non-negligible minority of around $5\% - 20\%$ of GRB light-curves show such a precursor. These precursors are characterized by peak flux that is smaller than a third of the main pulse, that  typically occur after a few 10's of seconds from the trigger, but with some cases reaching up to several 100's of seconds. In many cases, the precursors and the main pulses are separated by a distinct quiescent period. Similar figure, $9\%$ of bursts in the GBM catalogue showing evidence for a precursor was reported by \citet{Coppin+20}. We, therefore, raise the possibility that the main emission in these  burst light-curves are due to such a CBM emission phase, and not due to the traditional prompt emission phase.

The wind-blown bubbles have a complex density structure, basically composed of four distinct regimes \citep{CMW75, Weaver+77}: (a) unshocked stellar wind; (b) shocked stellar wind; (c) shocked interstellar medium (ISM) and (d) unshocked ISM. A forward propagating shock separates regions (c) and (d) (the unshocked and shocked ISM), a reverse shock separates regions (a) and (b), and a contact discontinuity (CD) separates regions (b) - the shocked stellar wind from region (c) - the shocked ISM. 

Assuming that the WR star is the progenitor of a GRB, this complex density structure is expected to affect the dynamics of the propagating relativistic blast wave. This scenario was studied by several authors \citep{PW06, NG07, VanEerten+09, MG11}, focusing on the interaction between the GRB blast wave and the wind termination (reverse) shock. These authors concluded that although some observed signal is expected as a result of this interaction, it is not expected to be significant.  We point out though, that the size of the bubble considered in these works, $\sim 10$ pc, implied that the predicted signal was at late times; during the afterglow phase of GRBs. While this size is typical for many systems, as pointed out above, it is much larger than what is inferred from observations of some nearby WR stars \citep{ Sirianni+98, Cohen+05}. Thus, for $\lesssim$~pc scale nebulae, the resulting signal may be different than predicted at earlier works, and occur at much earlier times.  

As we show here, although the interaction of the GRB blast wave and the wind termination shock may not lead to a strong observed signal, such a signal is expected as a result of the GRB blast wave interaction with the contact discontinuity (CD) that separates the shocked wind and the shock ISM.  The resulting light-curves from such a scenario should therefore have three different episodes: the triggering signal (resulting from the explosion that leads to the ejection of a relativistic jet), a quiescent period, followed by a bright, long flare of synchrotron origin that occurs when the blast wave interacts with the CD. 

In this paper, we  explore such a wind bubble, or nebula, scenario in order to explain light curves with a precursor which is  followed by a strong synchrotron emission episode.
This paper is organized as follows. In \S \ref{sec:2} we introduce our assumption
and describe the general properties of the wind bubble.  In \S \ref{sec:3} we describe the interaction of the blast wave with the wind bubble and, in particular, with the contact discontinuity in \S \ref{sec:4}. We discuss and conclude in \S \ref{sec:6}.

\section {The circumburst environment in the "wind bubble" scenario}
\label{sec:2}

The structure of the "wind bubble" depends on the evolutionary stages of the progenitor star. Here we adopt the common assumption that prior to the explosion that leads to a GRB, the progenitor star undergoes a Wolf-Rayet stage \citep{GML96a, GLM96b}. Indeed, 
wind bubbles are observed causing distinct ring nebulae around massive stars in their Wolf-Rayet stage \citep[e.g., ][]{JH65,  Marston95, marston97, DAB13, Lau+22}.  

Such wind-blown bubbles are composed of low-density gas ($< 10^{-3} \mathrm{cm}^{-3}$)
with temperatures exceeding $10^7$~K \citep{TA11, TA18}.  The low density and hot 
gas may not be directly detected due its low X-ray emissivity.
However, in the mixing zone between  the hot bubble and the cooler outer material  the emission measure is expected to increase to detectable levels. Such a scenario is consistent with observations yielding upper limits to the density \citep[e.g., ][]{Chu2003, TG13}. Furthermore, we note that such tenious medium is consistent with several observations of GRB afterglows, which  indicate circumburst medium which have very low densities \citep{Piro2014, Ryde2022, Dereli22}, in some cases, even as low as $ 10^{-4}\, {\rm cm}^{-3}$ 
\citep{Oganesyan2023}. Finally, the size of the cavity inside the Wolf-Rayet ring nebula depends on the metalicity of the  environment into which the wind emerges, with higher metalicity typically associated with a larger nebulae \citep[e.g., ][]{RR2001}. Therefore, one may consider the sizes that are measured locally in our own galaxy to serve as upper limits of the sizes that are expected in the GRB host galaxies.

We adopt here the following characteristics in describing the circumburst ring nebula:
\begin{itemize}
    \item Size: $\lesssim$1~pc. \citep[e.g., ][]{SB10, Sirianni+98, Cohen+05}.
    \item Characteristic density inside the cavity: $n \sim 10^{-2}~{\rm cm^{-3}}$ \citep[e.g., ][]{Chu2003, TG13}.
    \item Temperature $T \sim 10^7$~K \citep[e.g., ][]{Chu2003, TG13}. 
    \item To these we add an observational time of the GRB flare at $\sim 100$~s from the precursor, as well as typical flare duration of a few tens of s. \citep[e.g., ][]{Zhang+18, Sharma+19}.
\end{itemize}

The physical structure of the wind bubble was derived by \citet{CMW75}. The entire structure depends on only four parameters: (i) The mass loss rate, which we take as $\dot M = 10^{-9} {\dot M}_{-9} ~{\rm M_\odot/year}$. This value is lower than the fiducial value often taken in the literature of $10^{-5}~ {\rm M_\odot/year}$, and is chosen here being consistent with the observational data as given above and further discussed below, in particular the cavity size of $\sim 1$~pc, as well as the flaring GRB time of $\sim 100$~s. Further, this value is consistent  with  the estimates of \citet{Piro2014}. (ii) The stellar wind velocity, which is taken as $v_w = 10^8 ~v_{w,8}~{\rm cm~s^{-1}}$. As we show below, this value determines the temperature of the cavity, and therefore is fixed once this temperature is measured. (iii) The time the star emits the wind, which is taken as $t_\star = 10^4~t_{\star,4}$~yr. This value mainly affects the cavity size. This may be the most uncertain value, as the wind properties just prior to the final stellar explosion are least constrained. Finally, (iv) The ambient ISM density is taken as $n_{\rm ISM} = 10~ n_{0,1}~{\rm cm^{-3}}$. The value of $n_{\rm ISM}$ is related to the density outside the ring nebula, which itself can be part of a superbubble or an interstellar bubble \citep{Chu2016, Vam+16}. However, the value has relatively small influence on the observed signal, as we show below.

\begin{figure}[h]
    \centering
    \includegraphics[width=\linewidth]{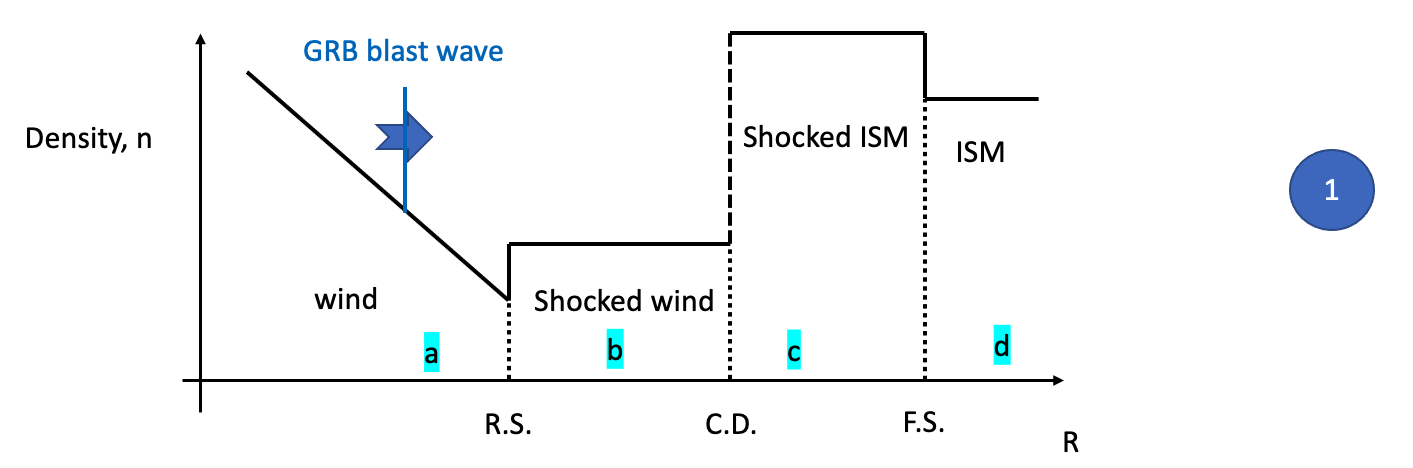}
    \caption{Cartoon showing the density profile of a wind bubble as a function of the distance, $R$,  from the central star. The four distinct regimes, marked (a)--(d) are separated by the forward and reverse shocks (F.S and R.S.), as well as the contact discontinuity (C.D.).}
    \label{fig:1}
\end{figure}

The wind bubble consists of four distinct regions. (a) A hypersonic stellar wind, characterized by a density decrease, $n(r) = \dot M/(4 \pi m_p r^2 v_w) \propto r^{-2}$ (for steady wind flowing at constant velocity). (b) A hot, almost isobaric region consisting of shocked stellar wind mixed with a small fraction of interstellar gas. (c) A thin, dense shell containing most of the swept-up interstellar gas; and (d) the ambient interstellar gas. In Figure \ref{fig:1}, we present a cartoon showing these four distinct regimes. 
In between regions (a) and (b) there is the wind reverse shock, also referred to as the wind termination shock (RS); in between regions (b) and (c) there is a contact discontinuity (CD); and in between regions (c) and (d) there is the wind forward shock (FS).

Neglecting the width of region (c) compared to region (b) (see below), and assuming that most of the energy in region (b) is in the form of thermal energy, it was shown by \citet{CMW75} that the forward shock (outer termination shock) radius is given by 
\beq
R_{\rm FS,W} = \left({125 \over 308 \pi} \right)^{1/5} \left({\dot M v_w^2 t_\star^3 \over \rho_{ISM}}\right)^{1/5} = 6.9 \times 10^{17}~\dot M_{-9}^{1/5} v_{w,8}^{2/5} t_{\star,4}^{3/5} n_{0,1}^{-1/5}~{\rm cm} 
\label{eq:R_FS}.
\eeq
The density in region (c) is approximated as $\rho_c \simeq 4 \rho_{ISM}$, valid for a strong, adiabatic shock. Then the width in region (c) is calculated by comparing the total ISM mass swept to radius $R_{\rm FS,W}$, $M_{ISM} \simeq (4\pi/3) R_{\rm FS,W}^3 \rho_{ISM}$ to the mass in region (c), $\approx 4 \pi R_{\rm FS,W}^2 \Delta R_c \rho_c$ giving $\Delta R_c = R_{\rm FS,W}/12$.

The pressure in regions (b) and (c) is $P_b = P_c = (2/3) u_b$ (assuming a monoatomic gas), where $u_b$ is the energy density in region (b), giving
\beq
P_b = P_c =  {7 \over 25} \left({125 \over 308 \pi} \right)^{2/5} \rho_{ISM} \left({\dot M v_w^2 \over  t_\star^2 \rho_{ISM}}\right)^{2/5} = 2.2 \times 10^{-11}~\dot M_{-9}^{2/5} v_{w,8}^{4/5} t_{\star,4}^{-4/5} n_{0,1}^{3/5}~{\rm dynes~cm^{-2}}. 
\label{eq:P_b}
\eeq
Using the strong reverse shock assumption, $P_b \gg P_a$, \citet{Weaver+77} calculated the radius of the reverse shock, by equating the ram pressure in the upstream region (a), $\rho_a(R_{\rm RS,W}) v_w^2$ to the pressure downstream, $P_b + \rho_b v_b^2$. This gives  
\beq
R_{\rm RS,W} = \left({3 \over 4} {\dot M v_w \over 4 \pi P_b}\right)^{1/2} = 1.3 \times 10^{17}~\dot M_{-9}^{3/10} v_{w,8}^{1/10} t_{\star,4}^{2/5} n_{0,1}^{-3/10}~{\rm cm}.  
\label{eq:R_RS}
\eeq
The number density of particles in region (a) (the unshocked wind) at $R_{\rm RS,W}$ is 
\beq
n_a(R_{\rm RS,W}) = {\dot M \over 4 \pi m_p R_{\rm RS,W}^2 v_w} = 1.8 \times 10^{-3} ~\dot M_{-9}^{2/5} v_{w,8}^{-6/5} t_{\star,4}^{-4/5} n_{0,1}^{3/5}~{\rm cm^{-3}}.  
\label{eq:n_a}
\eeq
The density in region (b) is assumed constant, and equal to $n_b \simeq 4 n_A(R_{\rm RS,W})$. The temperature in region (b) is calculated using the equation of state, $T_b = P_b/ k_B n_b$,
\beq
T_b = 2.27 \times 10^7~v_{w,8}^{2}~ \deg.  
\label{eq:T_b}
\eeq
One therefore finds that the temperature in region (b), namely the shocked wind regime, depends solely on the wind velocity. Thus, as stated above, a measurement of an observed temperature of $\sim 10^7~\deg$, provides a strong observational constraint, necessitating a wind velocity of the order of $10^8~{\rm cm~s^{-1}}$. Furthermore, note the strong dependence of the temperature on the wind velocity.

From Equation (\ref{eq:R_FS}) it is deduced that an observable bubble size of $\lesssim$~pc, implies that the wind duration cannot be much longer than $10^4$~yr, due to the very weak dependence of the cavity size on the mass loss rate, $\dot M$ and the ambient density, $n_0$. As the wind veocity is set by the cavity temperature, the only significant dependence of the cavity size is on the wind duration.

\section {The GRB blast wave as it crosses the wind bubble}
\label{sec:3}

At the end of its lifetime the progenitor star at the center of this wind bubble explodes, producing, by assumption a GRB. This explosion leads to the ejection of a blast wave (jet) that quickly becomes relativistic, being accelerated either by photon (radiative) pressure or by magnetic energy dissipation (magnetic reconnection). This relativistic blast wave then propagates through the complex wind bubble environment. As we show here, interaction of this expanding blast wave with the wind bubble environment can lead to an observed signal. A broad-band brightening is expected to occur and peak on an observed time scale of several tens of seconds following the stellar explosion. The origin of this brightening is the encounter of the blast wave with the contact discontinuity that separates regions (b) and (c). 

Material ejected from the star is distributed in regions (a) and (b), while regions (c) and (d) are composed of interstellar material.
The total rest mass energy in regions (a) + (b) is $E_{RM} = \dot M c^2 t_\star = 1.8 \times 10^{49}~{\dot M_{-9} t_{\star,4}}$~erg. This is several orders of magnitude lower than the isotropically equivalent energy released in a strong GRB explosion, $E_{GRB} \approx 10^{53}$~erg. \footnote{We point out, that this result is different than the assumptions used in the work by \citet{PW06}. In that work, a larger mass loss rate and longer wind emission time than considered here were assumed, resulting in the opposite conclusion.} This result implies that the blast wave will cross both regions (a) and (b) while still relativistic. 

Following the coasting phase, the blast wave will collect material from region (a), and the flow will eventually  become self-similar. As the shock wave reaches the reverse shock, its Lorentz factor is given by 
\beq
\Gamma(r=R_{\rm RS,W}) = \min \left\{ \Gamma_0, \Gamma_{BM}(r=R_{\rm RS,W})\right\},
\label{eq:Gamma_0}
 \eeq
where $\Gamma_0$ is the maximum Lorentz factor achieved from the explosion (i.e., during the coasting phase of the blast wave evolution), and $\Gamma_{BM}(r=R_{\rm RS,W})$ is the evolved Lorentz factor during the self-similar decaying phase, which is given by  
\citep{BM76}
\beq
\Gamma_{BM} (E, A; r=R_{\rm RS,W}) = \left({ 9 E \over 16 \pi A c^2 r}\right)^{1/2} = 1740~E_{53}^{1/2} \dot M_{-9}^{-13/20} v_{w,8}^{9/20} t_{\star,4}^{-1/5} n_{0,1}^{3/20}.
\label{eq:Gamma_BM0}
\eeq
Here, $A \equiv (\dot M/4 \pi v_w)$. 

The observed time at which the blast wave appears to reach the wind reverse shock is 
\beq
t^{ob.}({\rm RS,W}) \simeq (1+z) {R_{RS,W} \over 2 \Gamma^2(R_{RS,W}) c} = 1.4~\left({1+z \over 2}\right)~E_{53}^{-1}   \dot M_{-9}^{8/5} v_{w,8}^{-4/5} t_{\star,4}^{4/5} n_{0,1}^{-3/5}~{\rm s},
\label{eq:t0}
\eeq
namely, of the order of a second. In the calculation presented in Equation \ref{eq:t0} we assume the value of the Lorentz factor as given by the self-similar blast wave solution (Equation \ref{eq:Gamma_BM0}). As this value can be higher than $10^3$, It is more likely that the coasting Lorentz factor may be lower than that, of an order of $10^2- 10^3$ s is referred to in many GRBs. In such a case, the observed time scale for the blast wave to reach the reverse shock is longer than calculated above. However, as long as the initial GRB blast wave's Lorentz factor is a few hundreds, this time will not exceed a few seconds. This means that the emission associated with this interaction of the GRB blast wave with the wind termination (reverse) shock may be associated with the observed GRB precursor. Alternatively,  such a short time scale may leave little to no observed signal, as any signal, if exists, will overlap with the prompt emission phase, and will be identified as part of it.

The total mass swept by the shock in region (a) is $M = \int 4 \pi r^2 \rho(r) dr = \dot M R_{\rm RS,W}/ v_w$. This is much smaller than the total mass in regions (a) + (b) together, which is  $\dot M t_\star$. The ratio of the mass in region (a) to the total mass in regions (a)+(b) is therefore $R_{\rm RS,W}/v_w t_\star = 4.0 \times 10^{-3}~\dot M_{-9}^{3/10} v_{w,8}^{-9/10} t_{\star,4}^{-3/5} n_{0,1}^{-3/10}$, which is much smaller than unity. One  may therefore neglect the amount of swept up mass in region (a) relative to region (b). As a consequence, after a short transition phase that will occur once the GRB shock wave crosses the wind reverse shock, its motion will resume as being self similar.

As was shown by \citet{PW06}, once the blast wave encounters the reverse shock, it splits into two: a newly formed forward shock, whose initial Lorentz factor is $\Gamma (r=R_{\rm RS,W}+) = 0.725 \Gamma(r=R_{\rm RS,W}-)$, while the newly formed reverse shock moves at $0.43 \Gamma(r=R_{\rm RS,W}-)$. Here, $\Gamma(r=R_{\rm RS,W}-)$ is the Lorentz factor of the GRB blast wave just before encountering the wind termination shock, and is given by Equation \ref{eq:Gamma_0}. This newly formed reverse shock completes crossing the plasma at radius $1.06 \times R_{\rm RS,W}$, after which it disappears. The crossing causes a short re-brightening, on a short time scale of $0.06 R_{\rm RS,W}/(0.725 \Gamma(r=R_{\rm RS,W}-)^2/c \simeq 0.2$~s (see \citet{PW06} for details), after which only the forward shock will remain, and continue to propagate into region (b) in a self-similar pattern. Such a short re-brightening may be seen as an early pulse as part of the precursor phase.

\begin{figure}[h]
    \centering
    \includegraphics[width=\linewidth]{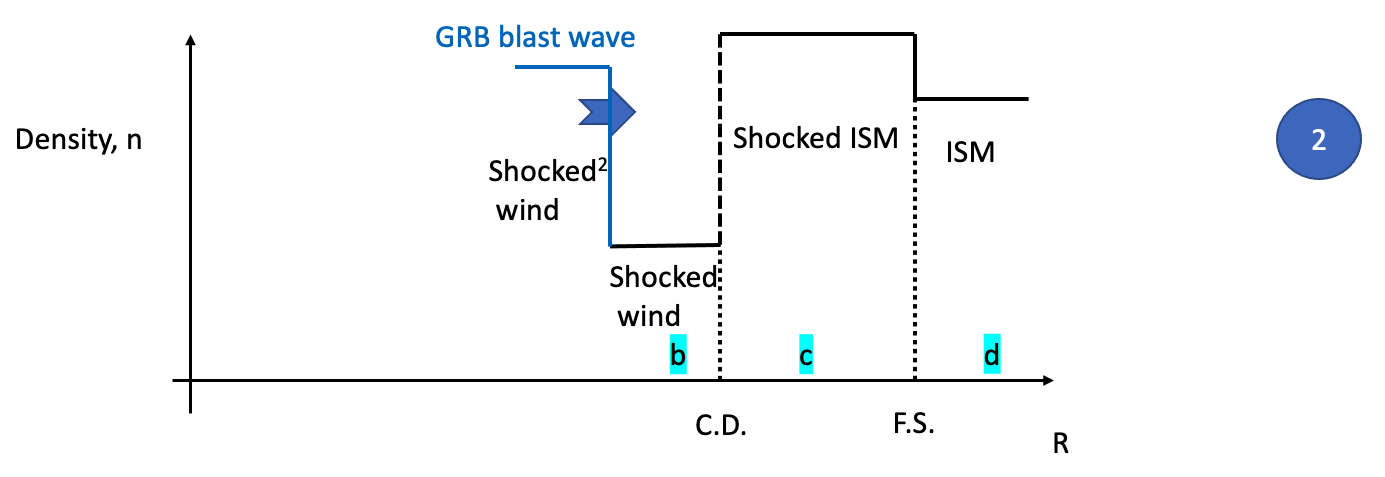}
    \caption{Similar to Fig. \ref{fig:1}, now showing the density profile after the blast wave crosses the wind's original reverse shock (R.S. in Fig.\ref{fig:1}). Behind the GRB blast wave, is material that was shocked twice: first, by the wind termination shock, and second by the GRB blast wave itself. Hence it is denoted in the cartoon as "shocked$^2$ wind".}
    \label{fig:2}
\end{figure}

Since the amount of material in region (a) is negligible relative to region (b), it is safe to assuming that after crossing the wind termination shock, the GRB blast wave continues to propagate in region (b) in a self-similar way, neglecting the short transition period. A cartoon  illustrating the blast wave propagation at this epoch is shown in Figure \ref{fig:2}. The blast wave Lorentz factor is given by the \citet{BM76} solution, using the prescription derived for an explosion into a medium having a constant density, as is expected by the shocked wind medium in region (b). When reaching the radius of the contact discontinuity, the blast wave Lorentz factor is equal to
\beq
\Gamma (E, n; r=R_{CD}) \equiv \Gamma_4 = \left({ 17 E \over 16 \pi n_b m_p c^2 R_{\rm CD}^3}\right)^{1/2} = 110~ E_{53}^{1/2} \dot M_{-9}^{-1/2} t_{\star,4}^{-1/2}
\label{eq:Gamma4}
\eeq
Here, we used $R_{\rm CD} = (11/12) R_{\rm FS,W}$, as explained above. We denote this Lorentz factor as "$\Gamma_4$", to simplify the discussion in the next section. 
The observed time at which the blast wave reaches the contact discontinuity is 
\beq
t^{ob.} = (1+z) {R_{\rm CD} \over 4 \Gamma^2(r) c} = 825~ \left({1+z \over 2}\right) ~E_{53}^{-1}  \dot M_{-9}^{6/5} v_{w,8}^{2/5} t_{\star,4}^{8/5} n_{0,1}^{-1/5}~{\rm s}.
\label{eq:t_ob.}
\eeq

\section{Interaction of the GRB shock wave with the contact discontinuity}
\label{sec:4}

On the external side of the contact discontinuity, region (c), the density is $n_c = 4 n_0 = 40~n_{0,1}~{{\mathrm{cm}}^{-3}}$ (to be compared with $4 \times 10^{-3}~{{\mathrm{cm}}^{-3}}$ in region (b), see below Equation (\ref{eq:n_a})). This means that the density jump is of factor $\sim 10^4$ for the parameters chosen, while the pressures and energy densities are the same. This can be understood, as the temperature in region (c) is colder by the same factor than that of region (b).  

\begin{figure}[h]
    \centering
    \includegraphics[width=\linewidth]{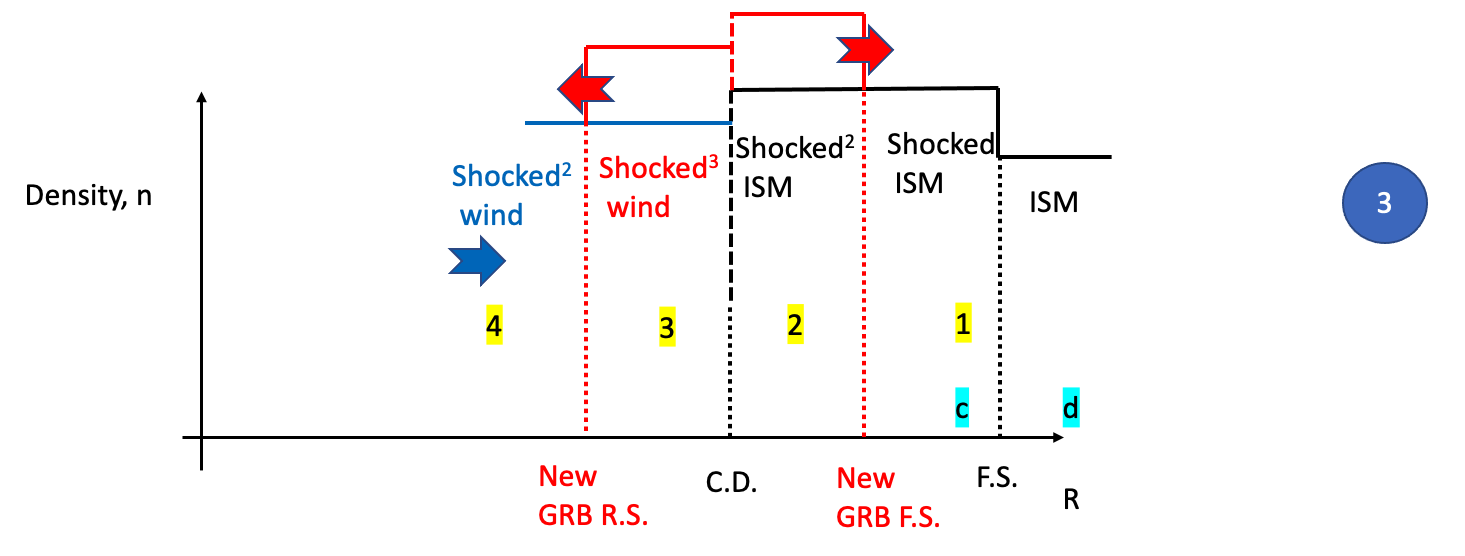}
    \caption{Similar to Fig. \ref{fig:1}, now showing the density profile after the blast wave crosses the contact discontinuity (C.D.). Region (1) was not yet reached by the newly formed forward shock, and is thus identical to previous region (c). Region (2) is composed of material from region (c), shocked by the newly formed forward shock. Region (3) is composed of wind material from the progenitor, first shocked by the wind termination shock (region (b) above), then by the GRB blast wave, and a third time by the newly formed reverse shock, hence is denoted by "shocked$^3$ wind". Region (4) is composed of the same material, before being reached by the newly formed reverse shock.}
    \label{fig:3}
\end{figure}

Once the propagating GRB blast wave reaches the contact discontinuity, it is split again into two: a forward shock that continues to propagate into region (c), and a reverse shocks. Similar to the crossing of the wind termination shock, the system becomes a 4-region system,  composed at this stage of the following regions.
(1) region (1) is what was referred to earlier as region (c), composed of the shocked ISM material that was not reached by the newly formed forward shock. Region (2) is composed of material from region (c), that was  shocked by the newly formed forward shock. Region (3) is composed of material  from region (b), shocked once by the GRB blast wave, and then again by the newly formed reverse shock. Note that since material in region (b) is composed of material that was earlier shocked by the wind termination shock, this material in fact was shocked three times (see cartoon in Figure \ref{fig:3}).
Region (4) is composed of material from region (b) that was shocked once by the GRB blast wave as it crossed it, but had not yet been reached by the newly formed reverse shock. It is thus marked in figure \ref{fig:3}) as "Shocked$^2$ wind". Figure \ref{fig:3} shows a cartoon of the four different regimes. 

Region (4) contains of material that cannot be considered cold, since it was already shocked by the GRB blast wave. region (1) on the other hand can be considered cold, since the shocked ISM in region (c) is still cold. 
This system was studied by \citet{Peer+17} and we adopt here the results derived by their equations 7-10, which are valid in the relativistic case. 

The enthalpy in each of the four regions is denoted by $\omega_i = e_i + p_i$. Here, $e_i = u_i + n_i m_p c^2$
is the energy density (including rest mass), $u_i$ is the thermal energy density (without the rest mass), and $p_i = (\hat \gamma -1) (e_i - n_i)$ is the pressure in region $i$, having density $n_i$. The adiabatic index is denoted by $\hat \gamma_i$, that assumes value $\hat \gamma_i = 4/3$ valid in the relativistic case.

The observed signal is expected from particles heated by the newly formed forward and reverse shocks, namely in regions (2) and (3). In order to calculate the plasma properties in these regions, one must use the properties of the surrounding plasma, i.e., in regions (1) [previously, (c) - the shocked ISM] and (4), which is the shocked wind, that was shocked again by the GRB blast wave. 

Region (1) is cold, namely $u_i \ll n_i m_p c^2$. This can be seen by using Equation \ref{eq:P_b}, which give $u_1 = (3/2) P_c = 3.3 \times 10^{-11}~{\rm erg~cm^{-3}}$, while $n_1 m_p c^2 = n_c m_p c^2 = 0.06~n_{0,1} ~{\rm erg~cm^{-3}}$, namely many orders of magnitude higher. Thus, the enthalpy in this region is 
\beq
\omega_1 \approx n_1 m_p c^2 = 0.06~n_{0,1} ~{\rm erg~cm^{-3}}.
\label{eq:omega1}
\eeq

Region (4) is composed of material from region (b), shocked by the GRB blast wave. The shock jump conditions in region (b) imply (for relativistic shock) that $u_4 = 4 \Gamma^2 n_b m_p c^2$, which is the dominant term in the enthalpy calculation, assuming that the GRB blast wave Lorentz factor is $\Gamma \gg 1$. Using Equations \ref{eq:Gamma4} and \ref{eq:n_a}, this gives (initially after the formation of the reverse shock)
\beq
\omega_4 = 0.54 E_{53}  \dot M_{-9}^{-3/5} v_{w,8}^{-6/5} t_{\star,4}^{-9/5} n_{0,1}^{3/5}.
\label{eq:omega4}
\eeq
The Lorentz factor of the newly formed shocked regions (2) and (3) is obtained using Equation (8) of \citet{Peer+17}, 
\beq
\Gamma_2 = \Gamma_3 \simeq \sqrt{\Gamma_4 \over 2} \left( \omega_4 \over \omega_1 \right)^{1/4} = \Gamma_4  \left( n_b \over n_c \right)^{1/4} = 13 E_{53}^{1/2}  \dot M_{-9}^{-2/5} v_{w,8}^{-3/10} t_{\star,4}^{-7/10} n_{0,1}^{-1/10}.
\label{eq:Gamma2}
\eeq
In between regions (1) and (2) is the forward shock, 
whose Lorentz factor is $\Gamma_{FS} \simeq \sqrt{2} \Gamma_2$;
and in between regions (3) and (4) is the reverse shock, whose Lorentz factor is
$\Gamma_{RS} \simeq \Gamma_3/\sqrt{2}$.

\subsection{Radiation from the shocked regions}

Being far from the central engine, it is safe to assume that the emission is in the optically thin region. One may therefore safely consider synchrotron emission as the main source of radiation. Using the standard assumptions that the shock waves accelerate electrons to a power law distribution with index $p$, the characteristic spectral breaks depend on only three ingredients: the bulk Lorentz factor, $\Gamma_2 = \Gamma_3$ as calculated in Equation \ref{eq:Gamma2} above; the characteristic electrons Lorentz factor, $\gamma_m$ (as measured in the comoving frame); and the magnetic field, $B$. I order to determine $\gamma_m$ and $B$, one first need to evaluate the plasma conditions in the emitting regions (2) and (3).  

The energy density in the newly shocked regions is given by Equation 9 of \citet{Peer+17}, 
\beq
\ba{lcl}
e_2 = e_3 & = & 2 \Gamma_4 \sqrt{\omega_1 \omega_4} 
 =  4 \Gamma_4^2 \sqrt{n_b n_c} m_p c^2 \\ & = & 41 E_{53}  \dot M_{-9}^{-4/5} v_{w,8}^{-3/5} t_{\star,4}^{-7/5} n_{0,1}^{4/5}~{\rm erg~cm^{-3}}.
 \label{eq:energy_density}
\ea
\eeq
which is of course equal in both regions. 
The energy per particle, in region (2) is: 
\beq
{e_2 \over n_2} = \sqrt{\Gamma_4 \over 2} {\omega_1^{3/4} \omega_4^{1/4} \over n_1} = 1.95 \times 10^{-2} E_{53}^{1/4}  \dot M_{-8}^{-2/5} v_{w,8}^{-3/10} t_{\star,4}^{-7/10} n_{0,2}^{-1/10}~ {\rm erg}
\label{eq:e_n2}
\eeq
and in region (3):
\beq
{e_3 \over n_3} = \sqrt{\Gamma_4 \over 2} {\omega_1^{1/4} \omega_4^{3/4} \over n_4} = 0.73 E_{53}^{-1/4}  \dot M_{-9}^{-3/5} v_{w,8}^{3/10} t_{\star,4}^{-3/10} n_{0,1}^{1/10}~ {\rm erg}
\label{eq:e_n3}
\eeq
where we used $n_4 = 4 \Gamma_4 n_b = 16 \Gamma_4 n_a(R_{\rm RS,W})$.

The energy density and energy per particle enables one to calculate the key ingredients of the synchrotron emission. Here we use the standard procedure, namely assume that uncertain fraction of the post-shock thermal energy, denoted by microphysical parameters, $\epsilon_B$ and $\epsilon_e$, is converted to magnetic field and used in heating the electrons to a power law distribution.
  
Using this prescription, the magnetic field in both regions (2) and (3) is given by $B^2/8\pi = \epsilon_B e_2$, resulting in
\beq
B = 3.2~E_{53}^{1/2}  \dot M_{-9}^{-2/5} v_{w,8}^{-3/10} t_{\star,4}^{-7/10} n_{0,1}^{2/5} \epsilon_{B,-2}^{1/2}~{\rm G}
\label{eq:B}
\eeq
where we assumed $\epsilon_B = 0.01$ though this value is highly uncertain. 
The characteristic electron Lorentz factor is given by $\gamma_{el} m_e c^2 = \epsilon_e (e/n)$. Thus, in region (2) one finds 
\beq
\gamma_m (2) = 2.4 \times 10^3~E_{53}^{1/4}  \dot M_{-9}^{-2/5} v_{w,8}^{-3/10} t_{\star,4}^{-7/10} n_{0,1}^{-1/10} \epsilon_{e,-1},
\label{eq:gamma_m2}
\eeq
while in region (3), 
\beq
\gamma_m (3) = 9.0 \times 10^4~E_{53}^{1/4}  \dot M_{-9}^{-3/5} v_{w,8}^{3/10} t_{\star,4}^{-3/10} n_{0,1}^{1/10} \epsilon_{e,-1}.
\label{eq:gamma_m3}
\eeq

\subsection{Characteristic synchrotron frequencies}

Synchrotron emission is characterized by the well-known broken power law spectrum, with two important spectral breaks: one at the peak of the emission, $\nu_m$ and one at the cooling, $\nu_c$ (we neglect synchrotron self absorption, which occur at frequencies well below those observed). Given the calculations of $\Gamma_2$, $B$ and $\gamma_m$ above evaluating these frequencies is straight forward. 

The observed peak frequency is $\nu_m^{ob.} = 4.2 \times 10^6 B \gamma_{m}^2 \Gamma$~Hz, and is thus 
\beq
\nu_m^{ob.} (2) =  9.9 \times 10^{11} ~{\rm Hz} = 4.1~ E_{53}^{3/2}  \dot M_{-9}^{-8/5} v_{w,8}^{-6/5} t_{\star,4}^{-14/5} n_{0,1}^{1/10} \epsilon_{e,-1}^2 \epsilon_{B,-2}^{1/2}~  {\rm eV},
\label{nu_{m,2}}
\eeq
for emission from region (2). 
Emission from the shocked region (3), peaks at 
\beq
\nu_m^{ob.} (3) =  1.4 \times 10^{18} ~{\rm Hz} = 5.8  E_{53}^{3/2}  \dot M_{-9}^{-2} t_{\star,4}^{-2} n_{0,1}^{1/2} \epsilon_{e,-1}^2 \epsilon_{B,-2}^{1/2} ~{\rm keV}. 
\label{nu_{m,3}}
\eeq
This means that if the parameter space regime is not too far from the one considered here, most of the emission in the keV-MeV band is expected from material shocked by the reverse shock. Material shocked by the newly formed forward shock is expected to contribute mainly in the optical band.

The second important spectral break is at the cooling frequency. Following the GRB blast wave interaction with the contact discontinuity, the underlying assumption  is that the newly formed reverse and forward shock waves accelerate particles. These particles than radiatively cool, resulting in a characteristic spectral break. The available time for cooling  is the (comoving) dynamical time $\Delta t'= \Delta R/\Gamma c$. Here, $\Delta R$ is a characteristic length scale crossed by the shock. This comoving time is related to the observed time using $\Delta t^{ob.} = \Delta t'/\Gamma$. One thus finds  
\beq
\ba{lcl}
\nu_c^{ob.} & = & 2.26 \times 10^{45} {\Gamma^3 \over B^3 \Delta r^2} ~{\rm Hz} = {2.26 \times 10^{45} \over c^2 \Gamma B^3} {1 \over {\Delta t^{ob.}}^2} \\
& = & {24 \over {\Delta t^{ob.}}^2}~  E_{53}^{-2}  \dot M_{-9}^{8/5}  v_{w,8}^{6/5} t_{\star,4}^{28/10} n_{0,1}^{-11/10} \epsilon_{B,-2}^{-3/2}  ~ {\rm MeV}
\ea
\label{eq:nu_c}
\eeq
where $\Delta t^{ob.}$ is the time (in the observer's frame) measured from the blast wave interaction with the contact discontinuity, i.e., with the re-brightening of the signal caused by the formation of the forward and reverse shock waves.

\subsection{Relative contribution of forward and reverse shock emission}

The total number of radiating particles shocked by the reverse shock is 
\beq
N_3 \sim \dot M t_\star / m_p = 1.2 \times 10^{52} ~\dot M_{-9} t_{\star,4}.
\label{eq:N3}
\eeq
The total number of particles available to radiate from the forward shock is 
\beq
N_2 = {4 \pi \over 3} R_{FS,W}^3 n_{ISM} = 1.4 \times 10^{55}  ~ \dot M_{-9}^{3/5} v_{w,8}^{6/5} t_{\star,4}^{9/5} n_{0,1}^{2/5}. 
\label{eq:N2}
\eeq
These results imply that the total number of particles radiating from the forward shock is three orders of magnitude greater than that from the reverse shock. If all the particles are radiating, the ratio of fluxes from regions (3) [material shocked by the revere shock] and (2) [material shocked by the forward shock] is 
\beq
{P_{syn,3} \over P_{sin,2}} = {N_3 B_3^2 \gamma_m(3)^2 \over N_2 B_2^2 \gamma_m(2)^2} = 1.24 
\label{eq:flux_ratio1}
\eeq
where we used $B_3 = B_2$ as well as the characteristic Lorentz factors calculated in equations \ref{eq:gamma_m2} and \ref{eq:gamma_m3}. Note that all the parametric dependence disappeared, and therefore this ratio is universal. Furthermore, note that this ratio is calculated in the comoving frame, and an observer will see a different ratio - see below.

\subsection{Observed pulse duration}

The observed pulse duration is set by the time it takes the forward and reverse shock waves to cross the plasma. As we show here, the reverse shock crosses the plasma at a much shorter time scale than the forward shock. 

The reverse shock propagates into region (4), composed of the shocked stellar wind that was shocked again by the GRB blast wave. As the reverse chock starts its propagation, this region now becomes the upstream region of the reverse shock. Following the argument derived in \citet{PW06}, the width of this region is the original width of the shocked wind region (b), divided by $8 \xi \Gamma$. Here, $\xi$ is a factor of the order unity, that takes into consideration geometrical effects. The original width of region (b) is $R_{CD}- R_{RS} \approx R_{CD}$, and the Lorentz factor $\Gamma$ can be approximated as the Lorentz factor of the propagating GRB blast wave prior to its interaction with the contact discontinuity, which is given in Equation \ref{eq:Gamma4}. 

The velocity of the reverse shock was calculated in \citet{PW06} (their equation 6). The calculation is straight forward, using conservation of particle flux across the reverse shock,  $n_4 \Gamma_4' \beta_4' = n_3 \Gamma_3' \beta_3'$. Here, velocities are measured in the reverse shock frame, and densities are in the comoving frame (of each individual region). Lorentz transform to the lab frame, in which the reverse shock moves at velocity $\beta_{RS}$, gives 
$ \Gamma_4' \beta_4' = \Gamma_4 \Gamma_{RS} ( \beta_4 - \beta_{RS})$ from which one readily derives 
\beq
\beta_{RS} = {n_4 \Gamma_4 \beta_4 - n_3 \Gamma_3 \beta_3 \over n_4 \Gamma_4 - n_3 \Gamma_3}
\label{eq:beta_RS1}
\eeq
The density in the upstream region (4) is that of the shocked wind, shocked again by the GRB blast wave. It is thus given by 
$n_4 \simeq 4 \Gamma_4 n_b$ and similarly $n_3 = 4 \bar \Gamma_3 n_4$. Here, $\bar \Gamma_3$ is the Lorentz factor of region (3) as measured in the frame of region (4). For an adiabatic index $\hat \gamma = 4/3$, the density ratio is $n_3/n_4 = 4 \bar \Gamma_3$, and $\bar \Gamma_3  \Gamma_3 = \Gamma_4/2$ \citep[e.g.,][their equations 7 and 8]{Peer+17}.
These give a particularly simple result for the reverse shock velocity,
\beq
\beta_{RS} = {\Gamma_4 \beta_4 - 2  \Gamma_4 \beta_3 \over \Gamma_4 - 2 \Gamma_4} = 2 \beta_3 - \beta_4.
\label{eq:beta_RS}
\eeq
Note that this is the velocity as measured by an external observer. 

The time it takes the reverse shock to cross the plasma is calculated as follows. Once the GRB blast wave reaches the contact discontinuity, the region behind it (compressed original region (c), which is now region (4)) has width $\Delta R_4 = R_{CD} / (8 \xi \Gamma_4)$. The time it takes the reverse shock to complete crossing the shocked plasma region  $c t_1 = \Delta R_4/ (\beta_4 - \beta_{RS} )$. 
However, an observer at infinity will measure time $\Delta t^{ob.} (RS) = ( 1- \beta_4) t_1 \simeq \Delta R_4/ 2(\beta_4 - \beta_{RS}) c \Gamma_4^2$.
Using Equation \ref{eq:beta_RS}, one can approximate 
\beq
2 (\beta_4 - \beta_{RS}) \Gamma_{4}^2 = 4  \Gamma_{4}^2 ( \beta_4 - \beta_3 ) \simeq 2 \Gamma_4^2 \left( {1 \over \Gamma_3^2} -  {1 \over \Gamma_4^2} \right) = 2 \left[  {\Gamma_4^2 \over \Gamma_3^2} - 1 \right] \approx 2  {\Gamma_4^2 \over \Gamma_3^2}. 
\label{Eq:time_Calc1}
\eeq
Overall, one finds that the observed time for the reverse shock to cross the expanding plasma is 
\beq
\Delta t^{ob.} (RS) = (1+z) {R_{CD} \Gamma_3^2 \over 8 \xi \Gamma_4 2 c \Gamma_4^2} = (1+z) \left({11 \over 12} \right) {R_{FS} \Gamma_3^2 \over 16 \xi c \Gamma_4^3} = 310 \xi^{-1} \left({1 +z \over 2} \right) ~ E_{53}^{-1/2} \dot M_{-9}^{9/10} v_{w,8}^{-1/5} t_{\star,4}^{7/10}  n_{0,1}^{-2/5}~s. 
\label{eq:t_RS}
\eeq

As opposed to that, the observed time during which the forward shock crosses the shocked ISM region (region (1)) is much longer. The forward shock reaches the edge of the wind bubble, which is a distance $\Delta R_1 = R_{FS,W} - R_{CD} = R_{FS,W}/12$ away, at time $t_1 = \Delta R_1/\beta_{FS} c$. However, the observed time is the time difference between the detection of a photon emitted at the contact discontinuity at $t = 0$ and a photon emitted at $\Delta R_1$ at time $t_1$, which is 
 $\Delta t^{ob.} = (1-\beta_{FS}) t_1 \simeq {t_1 / (2 \Gamma_{FS}^2)}$. 
 Using $\Gamma_{FS}^2 = 2 \Gamma_2^2$, this gives the well-known result,
\beq
\Delta t^{ob.} (FS) = (1+z) \left({1 \over 12} \right) {R_{FS,W} \over 2 \beta_{FS} c \Gamma_{FS}^2} = 5650  \left({1 +z \over 2} \right) ~ E_{53}^{-1} \dot M_{-9} v_{w,8} t_{\star,4}^{2} ~s
\label{eq:t_FS}
\eeq
We thus find that the time it takes the reverse shock to cross the plasma is $\gtrsim 15$ times shorter than the time it takes the forward shock to reach the edge of the wind bubble. This means that the observed flux from the reverse shock is $\gtrsim 15*1.24 \approx 18$ times larger than the flux emitted from the forward shock, since the forward shock accelerates only 1/15 of the plasma by the time the reverse shock is observed to complete its crossing. One may therefore neglect emission from the forward shock when considering the contribution to the x- and $\gamma$- ray flux.

\section{Comparison with observations}
\label{sec:5}

A non-negligible fraction, of $\sim 10-20\%$ of GRBs
show evidence for a distinct precursor in their x-ray lightcurve \citep{Lazzati05, Zhu15, Coppin+20}. 
In these GRBs, the main emission episode, often identified as the "prompt" emission, is delayed relative to the precursor by a typical time scale of a few tens of seconds, with some cases delayed up to $\gtrsim 100$~s. The duration of the prompt phase is typically a few tens of seconds; typically, the duration of the main emission episode is comparable to that of the quiescence time between the precursor and the main peak, while the precursor is about an order of magnitude shorter in time \citep{Zhu15}. 
The spectra shows a clear non-thermal behaviour, which can therefore be fitted with a synchrotron emission \citep{Zhu15}. These characteristics are in line with the model proposed here, in which the main emission originates from the reverse shock that follows the interaction of the GRB blast wave with the contact discontinuity. 
In such a case, the main emission episode should not be identified as the prompt phase, but rather the CBM phase, since its origin is different.

As a specific example, we consider GRB160821A \citep{Sharma+19}, whose lightcurve and spectra were well analyzed \citep{Ryde2022}. This was a particularly bright burst, with energy flux of $2.86 \times 10^{-6} ~{\rm erg~cm^{-2}~s^{-1}}$ in the 10-1000 keV band. This makes the burst the third brightest GRB observed by
the {\it Fermi Gamma-ray Space Telescope} \citep{Atwood09} to date in terms of energy flux. This burst's lightcurve showed a short, weak precursor, followed by a quiescence period of 112~s, before the main emission episode took place. Once started, this episode lasted for 43~s \citep{Sharma+19}. Spectral analysis revealed that the main emission episode is well fitted by a synchrotron model, with peak frequency at $\nu_m^{ob.} \sim 5$~MeV anf a spectral cooling break at $\nu_c^{ob.} \simeq 100$~keV \citep{Ryde2022}.

As the redshift of this GRB is unknown, an exact value of the emitted energy is uncertain. For $z=1$, it will be $7 \times 10^{54}$~erg, while for $z=0.2$ the released energy is $8 \times 10^{53}$~erg. We may choose here a conservative value of $E = 10^{54}$~erg. Then use of equations (\ref{eq:t_ob.}) and 
(\ref{eq:t_RS}) implies that the main emission episode is expected at $\sim 83$~s, and its duration is $\sim 50$~s, governed by emission from the reverse shock (no redshift added here). Furthermore, if one assumes an equipartition value of the electrons's energy, namely $\epsilon_e = 0.33$, 
one finds, using equations (\ref{nu_{m,3}}) and (\ref{eq:nu_c}) a peak frequency at $\nu_m^{ob.}(3) \approx 2$~MeV and cooling break at 
$\nu_c^{ob.} \approx 240/(\Delta t^{ob.})^2$~keV. These values are surprisingly close to the values of these parameters inferred from observations.
An additional degree of freedom is the fraction of energy carried by the magnetic field ($\epsilon_B$). Values smaller (or higher) by the fiducial value of $10^{-2}$ taken here can be considered, resulting in some flexibility in the cooling break (and, to a lesser extend, in the peak energy).

\section{Discussion and Conclusions}
\label{sec:6}

There is ample evidence that the progenitors of long gamma-ray bursts (GRBs) are massive, Wolf-Rayet stars. These stars emit strong winds prior to their terminal explosion, producing a density structure known as "wind bubble" \citet{Weaver+77}, that is composed of four distinct regions: (i) the stellar wind; (ii) the shocked stellar wind; (iii) the shocked interstellar material (ISM) gas; and (iv) the unshocked ISM. Observations reveal that this wind bubble has a typical size of $\sim$~pc \citep{Cohen+05, TG13}. When the star explodes producing a GRB, the GRB jet interacts with this environment. Here, we consider a plausible scenario, in which the wind bubble is of $\lesssim$~1 pc size. We have shown above that in this scenario, the blast wave will encounter the wind termination shock on a very short time scale, of a few seconds or less (Eq. \ref{eq:t0}), and that the main energy dissipation mechanism will take place when the blast wave will encounter the contact discontinuity, separating the shocked stellar wind from the shocked interstellar medium. As the blast wave encounters this contact discontinuity separating, after about $\lesssim 100$~s observed quiescence period (Eq. \ref{eq:t_ob.}), the reverse shock can produce a bright flash, lasting a few tens of seconds (Eq. \ref{eq:t_RS}). This setup may thus provide a natural explanation to the brightening seen at this epoch in a substantial minority, $\sim 10$~\% of GRBs \citep{Lazzati05, Zhu15, Coppin+20}.

This scenario thus provides a natural explanation to a significant sub-sample of long GRBs. In fact, it helps solving one of the well known problems, that of radiation efficiency \citep{Peer15, Zhang18}. The nature of the energy dissipation that leads to the observed radiation. Kinetic models, such as the GRB "fireball" model considers internal shock waves within the expanding plasma as a mechanism of kinetic energy dissipation. However, a well known problem of this model is that of the low efficiency: typically, only a small fraction, of the order of a few \% of the kinetic energy is dissipated \citep{DM98, Spada+00}. This serves as a motivation to search for alternatives, such as models involving reconnection of magnetic field lines as a source of energy \citep{ZY11, SS14}. The scenario here suggests an elegant solution to this problem. As the GRB blast wave encounters the contact discontinuity, which is nearly static, a significant fraction of its kinetic energy is dissipated, leading to a strong signal.  

The characteristic time scales, both of the quiescence time prior to the main pulse, as well as the pulse duration itself, depends on the uncertain properties of the wind bubble (mainly its size and internal density), as well as the properties of the GRB blast wave. While there is a large uncertainty, we rely on measurements of wind bubbles seen in our Galaxy to estimate the size of the cavity. The values used for the mass loss rate as well as the typical wind ejection time are lower by several orders of magnitude than those typically found in the literature when considering Wolf-Rayet stars. However, they do match the observed cavity sizes of $\lesssim$~1 pc. Given the uncertainty that exists in the properties of the GRB progenitor stars in the last thousands to millions of years prior to their explosion, we believe that the parameters considered are plausible ones. 

Similarly, while the calculation itself is by large, one dimensional, when calculating the expected observed times we do consider emission from off-axis angles (known as the "high latitude" emission). These are taken into account by the pre-factors used in relating the observed time to the radius in which the interaction takes place ($t^{ob.} \propto R/\Gamma^2$. Details of the calculations appear in \citep{Waxman97}, as well as in the appendix of \citep{PW06}. While these pre-factors are of the order of a few, they imply that the expected ratio of the the delay time to the pulse duration can be of that order (i.e., of a few), as is indeed observed: e.g., in GRB160821A, the quiescence period between the precursor and the main pulse is 112~s, while the main pulse duration is 43~s; namely a ratio of less than 3. Such a ratio can well be explained by the different pre-factors.   
Since most precursor bursts indeed have such modest ratios \citep[e.g., ][]{Lazzati05, Burlon08, Hu14, Wang20, Deng24} the presented scenario, in its simplest form, can easily accommodate the observed properties. However, in a few cases the ratio is larger than a few, e.g. in GRBs 150330A and 160625B \citep{Deng24}. This might indicate that, in these cases, it is not the same blast wave giving rise both to the precursor and to the main burst event, with earlier central engine activity needed \citep[e.g., ][]{LP07}.

In some of the main pulses detected, light fluctuations ("spikes"), on a shorter time scale, of a few seconds were observed \cite{Zhu15, Sharma+19}. Such spikes can be accomodated within the framework of the model presented here, as the contact discontinuity is expected to be hydrodynamically unstable, similar to the situation in supernovae remnants  \citep{CBE92}. As a result, it will develop patches, which, when encountered by the GRB blast wave, may lead to the observed fluctuations in the lightcurve. Moreover, infrared observations of Wolf-Rayet stars reveal, in many cases,  clumpy structures of the nebula region \citep{Vam+16}.  This would further add to the possibility of fluctuations in the light curve on shorter time scales.

In fact, within this scenario, the detailed morphological shape of the light curve of the main emission will serve as a tomography of the immediate, complex density environment surrounding the progenitor star. Clumpy structures  \citep{Vam+16} and multiple rings within the nebula region \citep{Marston95} will lead to a variable, emitted intensity.  Interestingly, the same density environment could also be probed by infrared spectroscopy by identifying the complex absorption profiles in the infrared \citep[see, e.g., ][]{CT2010}.

The main pulse originates from the reverse shock that forms once the GRB blast wave reaches the contact discontinuity. However, at the same time, the newly formed forward shock continues to propagate towards the edge of the cavity. The signal from this forward shock is expected to be much weaker than that from the reverse shock. For the parameters chosen, the forward shock will reach the edge of the cavity on a time scale of $\sim$~1 hour (see Eq. \ref{eq:t_FS}, after which is will propagate into the interstellar material. As the forward shock generates the afterglow emission, a prediction of this model is a break seen in the afterglow emission at this time scale, of roughly an hour. After that period, the afterglow is expected to evolve according to the classical, \citet{BM76} solution. In fact, detection of such a break, when combined with the properties of the main peak, may serve as a new tool in studying the properties of the wind bubble. 

The expected spectra during the main pulse is expected to be dominated by sycnhrotron emission, with characteristic breaks that are given in Equations (\ref{nu_{m,3}}) and  (\ref{eq:nu_c}). The exact values of these breaks depend on the uncertain microphysical parameters ($\epsilon_e$ and $\epsilon_B$) that determine the post-shock thermal energy converted to electrons and magnetic field. Despite this uncertainty, these are believed to be universal parameters, hence their values are expected to be similar in the forward and reverse shock waves. Thus, the ratio of the power radiated from the forward and reverse shocks is expected to be universal.

To conclude, the wind bubble scenario presented here provides a natural scenario that can explain a significant minority of GRBs that show a precursor, followed by a quiescence period of several tens - hundreds of seconds before the main peak, which lasts a similar, though somewhat shorter duration of several tens of seconds.
In addition to be a natural outcome, this model has the advantage of high efficiency in the energy dissipation, producing the synchrotron emission. This phase is thus distinct from both the prompt and the afterglow phases, and should be better referred to as the CBM phase. 
If correct, studies of the CBM phase lightcurves and spectra could be a particularly useful tool in determining the shapes of wind bubbles, hence constrain the properties of GRB progenitor stars, just prior to their terminal explosion.

\section*{Acknowledgements} 

We thank The Swedish Foundation for International Cooperation
in Research and Higher Education (STINT) for support through the Internationalisation grant for scientific cooporation between Sweden and Israel (IB 2019-8160). We further acknowledge support from the Swedish National Space Agency (2021‐00180 and 2022‐00205).  AP acknowledge the support from the European Union (EU) via ERC consolidator grant 773062 (O.M.J.).

\bibliography{references}{}
\bibliographystyle{aasjournal}

\end{document}